\pdfoutput=1 
\documentclass{article}
\usepackage{dcase2020_techrep,amsmath,graphicx,url,times,booktabs, tabularx}
\usepackage{multirow}
\usepackage{xcolor}
\usepackage{comment}
\usepackage{amssymb}

\title{Deep Dense and Convolutional Autoencoders 
for Unsupervised Anomaly Detection in Machine Condition Sounds}

 \name{Alexandrine Ribeiro$^{1}$,
      Lu\'{i}s Miguel Matos$^{2}$,
      Pedro Jos\'{e} Pereira$^{2}$, 
      Eduardo C. Nunes$^{2}$,
     }
 \secondlinename{
       Andr\'{e} L. Ferreira$^{3}$,
       Paulo Cortez$^{2}$, 
       Andr\'{e} Pilastri$^{1}$
       }

 \address{$^1$ EPMQ - IT Engineering Maturity and Quality Lab, CCG ZGDV Institute, Guimar\~{a}es, Portugal,\\ 
        \{alexandrine.ribeiro, andre.pilastri\}@ccg.pt\\ 
         $2$ ALGORITMI Centre, Dep. Information Systems, University of Minho, Guimar\~{a}es, Portugal, \\  
         \{luis.matos, pedro.pereira, pcortez\}@dsi.uminho.pt\\ 
         \{b12176\}@algoritmi.uminho.pt\\
        $^3$ Bosch Car Multimedia, Portugal,\{Andre.Ferreira2\}@pt.bosch.com\\
    }

\begin{document}

\ninept
\maketitle

\begin{sloppy}

\begin{abstract}
This technical report describes two methods that were developed for Task 2 of the DCASE 2020 challenge. The challenge involves an unsupervised learning to detect anomalous sounds, thus only normal machine working condition samples are available during the training process. The two methods involve deep autoencoders, based on dense and convolutional architectures that use mel-spectogram processed sound features.
Experiments were held, using the six machine type datasets of the challenge. Overall, competitive results were achieved by the proposed dense and convolutional AE, outperforming the baseline challenge method.

\end{abstract}


\section{Introduction}
\label{sec:intro}

This work is motivated by a real-world task from the challenge on Detection and  Classification of  Acoustic Scenes and  Events (DCASE): unsupervised Anomalous Sound Detection (ASD). 
The DCASE challenge had its first edition in 2013 and three more editions from 2016 to 2019, with distinct learning tasks, ranging from acoustic scene classification to sound event detection.

In this technical report, we address the second task from the current DCASE edition (2020): Unsupervised Detection of Anomalous Sounds for Machine Condition Monitoring \cite{koizumi2020}.
The task aims to automatically detect, and as soon as possible, if a given machine is not working correctly by using only on the sound produced by the machine. Such an anomaly detection model is thus value for preventing future machine issues (e.g., equipment damage). 
The main challenge is to detect abnormal sounds using only standard working machine sound samples, assuming that the sounds produced by mechanical anomalies on the equipment are unknown. 
Although it may seem a binary classification problem (``normal'' or ``anomaly''), since the models can only be trained using data from one class, this task must be solved with an unsupervised learning anomaly detection approach.

For this task, a baseline system implementation was provided for comparison purposes \cite{koizumi2020}. The baseline consists of a dense Autoencoder (AE) with three layers, in  both  the encoder  and  decoder components, with  $128$  units, and a latent space with $8$ units, all with the ReLU activation function. 
In this paper, we propose two deep learning models, based on a Dense and Convolutional architectures fed with mel-spectograms,
which are further detailed in the next section.

\section{Methods}
\label{sec:methods}

\subsection{Datasets}
The data used for this task comprises parts of ToyADMOS \cite{koizumi2019toyadmos} and the MIMII Dataset \cite{Purohit_DCASE2019_01} consisting of the normal and anomalous operating sounds of six types of toy/real machines. This data was provided in two datasets (development and evaluation) for 6 different machine types: ToyCar, ToyConveyor, slider, pump, fan, and valve.
In the development dataset, each machine type has 4 different machines, except for ToyConveyor, which has only 3.
Moreover, normal and anomaly labels were provided for the test data, such that the anomaly detection performance could be estimated and the model could be tuned accordingly.
Regarding the evaluation data, it contains data for new machines in each machine type, both for model training and testing. Moreover, no labels are provided. A different number of approximately $10$ second Waveform Audio File (WAV) files is provided for each machine. Table~\ref{tab:data} summarizes the challenge datasets.
\begin{table}[!htb]
    \caption{Summary of provided datasets}  
    \centering
    \label{tab:data}
    \begin{tabular}{lcccc}
    \toprule
    \multirow{2}{*}{Machine Type} & \multirow{2}{*}{Mode} & \multirow{2}{*}{Machine ID} & \multicolumn{2}{c}{Audio Files} \\
    \cline{4-5}
    &  &  & Train & Test  \\
    \hline
    \multirow{7}{*}{ToyCar}
        & \multirow{4}{*}{Dev.}
            & 01 & 1000 & 614 \\
            & & 02 & 1000 & 615 \\
            & & 03 & 1000 & 615 \\
            & & 04 & 1000 & 615 \\
        \cline{2-5}
        & \multirow{3}{*}{Eval.}
            & 05 & 1000 & 515 \\
            & & 06 & 1000 & 515 \\
            & & 07 & 1000 & 515 \\
    \hline
    \multirow{6}{*}{ToyConveyor} 
        & \multirow{3}{*}{Dev.}
            & 01 & 1000 & 1200 \\
            & & 02 & 1000 & 1155 \\
            & & 03 & 1000 & 1154 \\
        \cline{2-5}
        & \multirow{3}{*}{Eval.}
            & 04 & 1000 & 555 \\ 
            & & 05 & 1000 & 555 \\
            & & 06 & 1000 & 555 \\
    \hline
    \multirow{7}{*}{fan}
        & \multirow{4}{*}{Dev.}
            & 00 & 911 & 507 \\
            & & 02 & 916 & 549 \\
            & & 04 & 933 & 448 \\
            & & 06 & 915 & 461 \\
        \cline{2-5}
        & \multirow{3}{*}{Eval.}
            & 01 & 934 & 426 \\
            & & 03 & 912 & 458 \\
            & & 05 & 1000 & 458 \\
    \hline
    \multirow{7}{*}{pump}
        & \multirow{4}{*}{Dev.}
            & 00 & 906 & 243 \\
            & & 02 & 905 & 211 \\
            & & 04 & 602 & 200 \\
            & & 06 & 936 & 202 \\
        \cline{2-5}
        & \multirow{3}{*}{Eval.}
            & 01 & 903 & 216 \\
            & & 03 & 606 & 213 \\
            & & 05 & 908 & 348 \\
    \hline
    \multirow{7}{*}{slider}
        & \multirow{4}{*}{Dev.}
            & 00 & 968 & 456 \\
            & & 02 & 968 & 367 \\
            & & 04 & 434 & 278 \\
            & & 06 & 434 & 189 \\
        \cline{2-5}
        & \multirow{3}{*}{Eval.}
            & 01 & 968 & 278 \\
            & & 03 & 968 & 278 \\
            & & 05 & 434 & 278 \\
    \hline
    \multirow{7}{*}{valve}
        & \multirow{4}{*}{Dev.}
            & 00 & 891 & 219 \\
            & & 02 & 608 & 220 \\
            & & 04 & 900 & 220 \\
            & & 06 & 892 & 220 \\
        \cline{2-5}
        & \multirow{3}{*}{Eval.}
            & 01 & 679 & 220 \\
            & & 03 & 863 & 220 \\
            & & 05 & 899 & 500 \\
    \bottomrule
    \end{tabular}
\end{table}

\subsection{Autoencoders}

The base learner is based on a AE, which has obtained good results in several studies \cite{provotar2019unsupervised_1, koizumi2018unsupervised_2, tagawa2015structured_3, kawaguchi2017can_4}. 
The AE is a specific artificial neural network in which the input is expected to be equal to the output and there are several hidden layers with fewer nodes than the number of inputs. The AE learning goal is to produce the same output for the same input, thus encoding and decoding the input signal via the hidden processing layers.

The encoder component of the AE maps the input vector (the features) into an hidden representation with a lower dimensional space, via a nonlinear transform.
Then, the decoder component attempts to reconstruct the reverse transform, from the hidden representation to the original input signal.
The difference between the original input vector and the AE output response is called the reconstruction error \cite{an2015variational_6}. 

In this challenge, the reconstruction error element is used to detect sound anomalies. Firstly, an AE is trained with only normal sound samples,
aiming to minimize the reconstruction error. 
The obtained model is assumed to be capable of compressing the input features, learning their most relevant relationships.
Secondly, the trained AE can be tested with unseen data. If the unseen data is similar to the trained patterns (related to the normal sounds), when
the AE should reproduce the new input with good accuracy. However, if the unseen data is anomalous,  the AE should not be able to reconstruct the input and
the error will be greater. Thus, the magnitude of the reconstruction error can be used to detect anomalies.
The proposed unsupervised anomaly detection approaches consist of simple AE networks. The systems were modeled to be generic, only changing the training data fed to the model, hence creating a generic model for anomaly detection in several machines. 

\subsection{Dense Autoencoder}
\label{ssec:systemx}

The first approach proposed, consists of a deep fully-connected AE (top of Figure~\ref{fig:networks}). 
After performing several preliminary experiments with different architectures (varying number of layers, hidden units, activation function and latent space dimensions), a final dense AE architecture was selected.
The encoder and decoder networks consist of four fully-connected layers with $512$ hidden units, followed by Batch Normalization 
and ReLU as the activation function. 
The bottleneck layer is set as one fully-connected layer with $8$ hidden units, resulting in a $8$-dimensional latent space.
\begin{figure}[!htb]
      \centering
      \centerline{\includegraphics[width=\columnwidth]{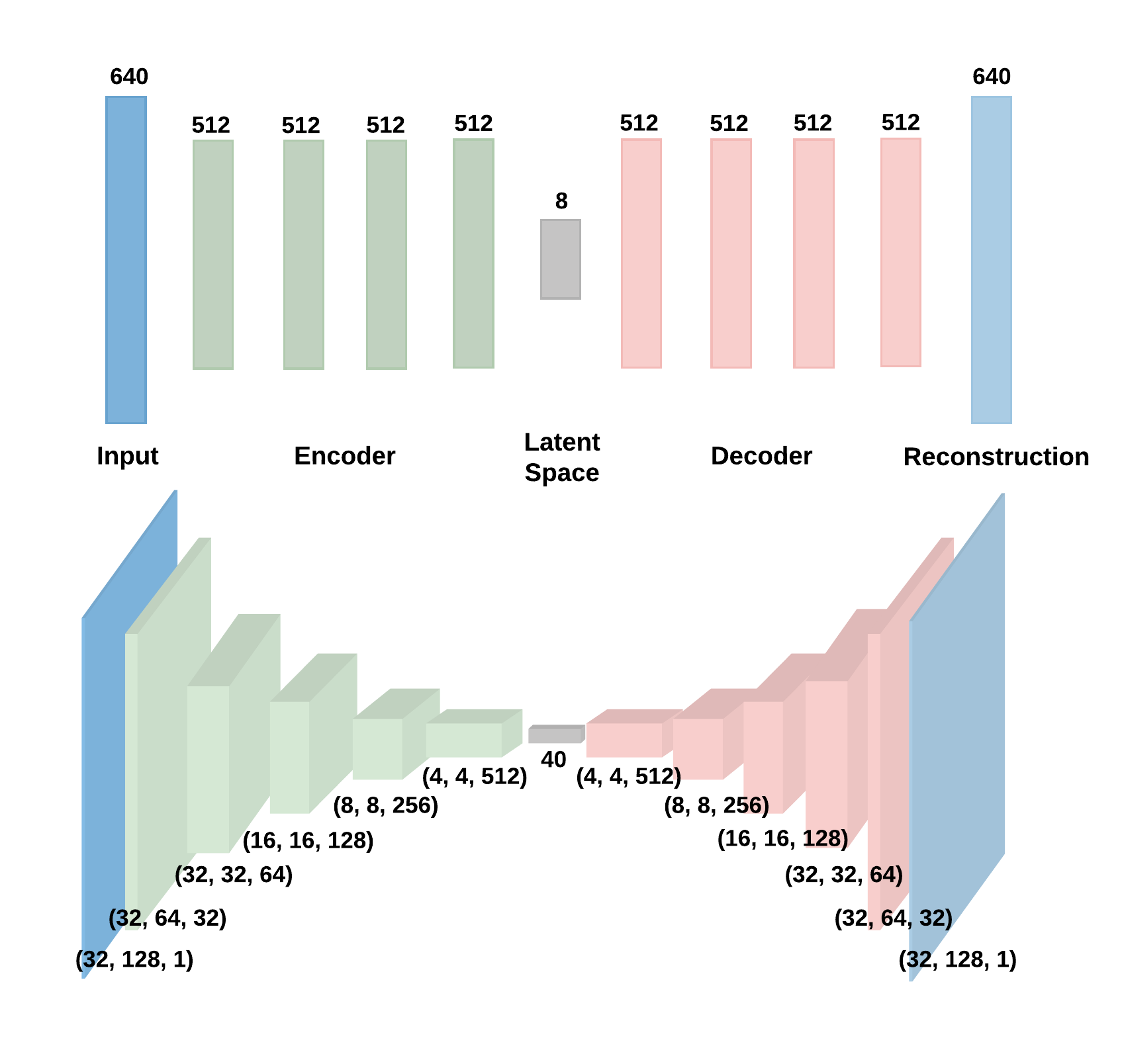}}
      \caption{Proposed AE Network architectures: Dense AE (top) and Convolutional AE (bottom).}
      \label{fig:networks}
    \end{figure}

\subsection{Convolutional Autoencoder}
\label{ssec:systemy}

Recently, Convolutional Neural Networks (CNNs) have been increasingly applied for audio processing tasks by using audio spectrograms as features \cite{li2017comparison, hershey2017cnn, chen2020novelty}. CNNs are an effective way to capture spatial information from multidimensional data being naturally suitable for 
exploring image-like time-frequency representations of audio, such as spectrograms.
The main goal of a CNN is to learn local structure in input data. Locality is a key property of CNNs. This is accomplished by convolutional filters that are applied to local regions of the previous layers to capture local patterns. Consequently, spatial features must be locally correlated. As time-frequency representations of audio are treated as images by CNN architectures, these features should be locally correlated in the sense of time and frequency.

The second approach proposed for the ASD task consists of a deep CNN  AE (shown in the bottom of Figure~\ref{fig:networks}).
Similarly to the dense AE network, preliminary experiments were used to adjust the CNN AE.
The encoder and decoder networks are comprised of convolutional layers with Batch Normalization and the ReLU activation function after each convolution.
The encoder network consists in a stack of five hidden layers with convolutional filters of $32$, $64$, $128$, $256$, and $512$, kernel sizes of $5$, $5$, $5$, $3$, and $3$, and strides of $(1,2)$, $(1,2)$, $(2,2)$, $(2,2)$, and $(2,2)$, respectively. 
The bottleneck consists of a convolutional layer with $40$ convolutional filters, reducing the encoder feature maps to a $40$-dimensional compressed representation of the input.  Regarding the decoder network, first a fully-connected layer inflates the latent space to the shape the last layer of the encoder, followed by five transposed convolutional layers that mirror of the encoder layers.

\subsection{Audio Features}
\label{ssec:features}

Regarding the feature engineering process, we have initially considered two main sound processing methods:
Mel Frequency Cepstral Coefficients (MFCCs) and Mel Frequency Energy Coefficients (MFECs).
MFCCs, which are derived from the mel-cepstrum representation of the audio, are one of the best knowns and most popular audio processing features \cite{sharma2020trends}. However, when computing MFCCs, a Discrete Cosine Transform (DCT) is applied to the logarithm of the filter bank outputs, resulting in decorrelated MFCC features. Therefore, they have the drawback of having non-local features, which makes them unsuitable for CNN processing.  
As such, in this work we explored a different feature for audio signal processing named MFECs, which are log-energies derived directly from the filter-banks energies. These are similar to MFCCs, however, they do not include the DCT operation. This feature provided good results in detecting different audio sounds and classification of sounds \cite{JAM2009,TorfiIND17}. Therefore, MFECs were selected as audio features for the proposed systems.

\section{EXPERIMENTS AND RESULTS}
\label{sec:results}
In this section, we describe our pipeline, including the feature preprocessing, model settings, hyperparameters and results obtained for both AE architectures.

\subsection{Features Extraction}
\label{ssec:extraction}

In the Dense Autoencoder system, audio data is buffered in fixed-length $1$ second intervals with a $50\%$ overlap. 
For each audio buffer obtained, the segment is then divided into $64$ ms analysis frames, with a $50\%$ overlap and $128$ MFECs extracted from the magnitude spectrum of each frame.  Then, a context window of size $5$ is used. Thus, $5$ frames are concatenated to form a $640$-dimensional input vector. 
This representation is depicted in Figure~\ref{fig:features_dense}.
\begin{figure}[!htb]
      \centering
      \centerline{\includegraphics[width=0.97\columnwidth]{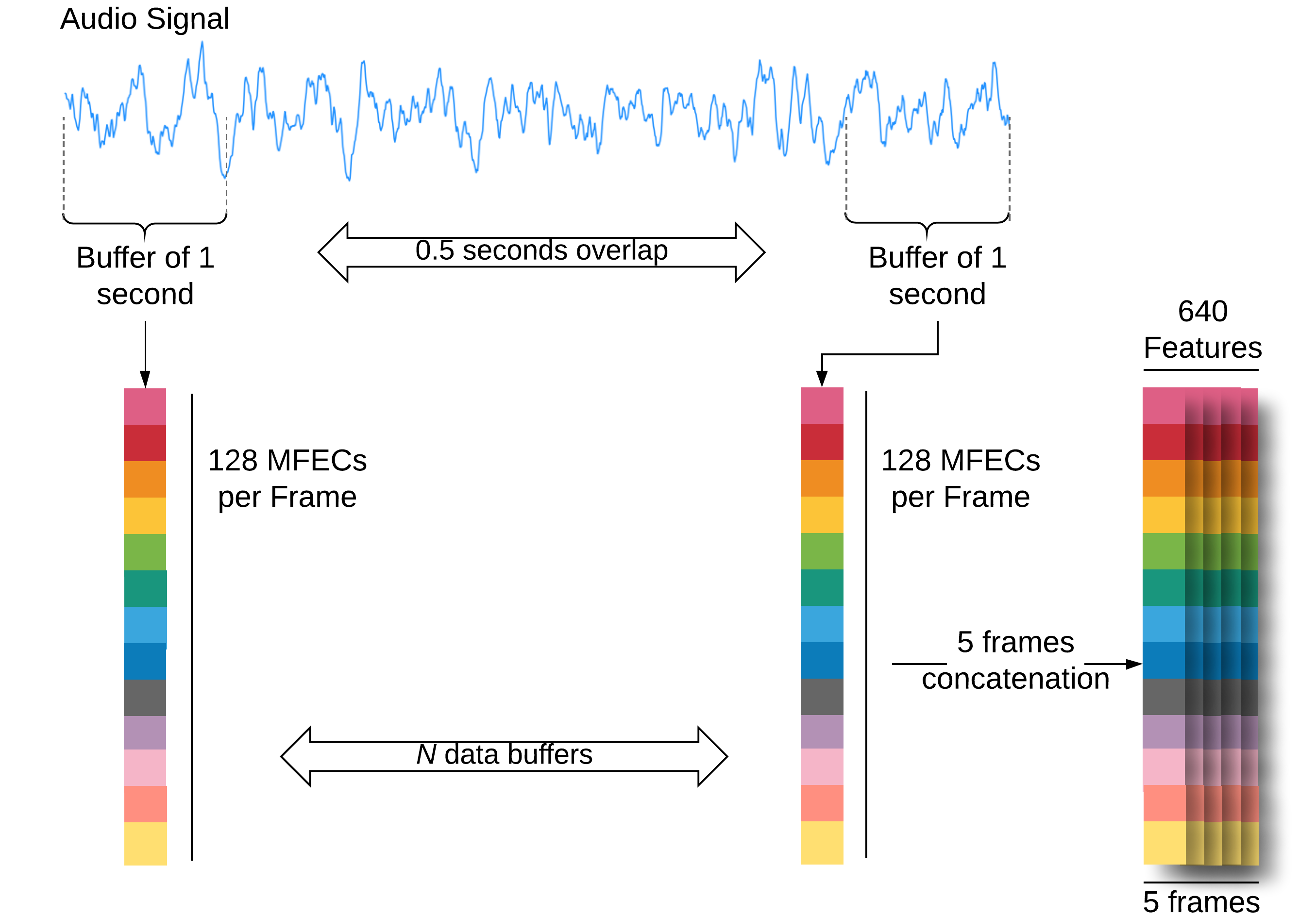}}
      \caption{Feature extraction procedure of the dense AE.}
      \label{fig:features_dense}
    \end{figure}

In the Convolutional Autoencoder system, for each audio, $128$ log mel-band energy features are extracted from the magnitude spectrum, considering $64$ ms analysis frames with $50\%$ overlap. 
Then, each feature is normalized to zero mean and unit standard deviation by using statistics from the training data.
Finally, the mel spectrogram is segmented about every second into $32$ column data with approximately $100$ ms of hop size.
This extraction procedure is shown in Figure~\ref{fig:features_conv}.
    \begin{figure}[!htb]
      \centering
      \centerline{\includegraphics[width=0.8\columnwidth]{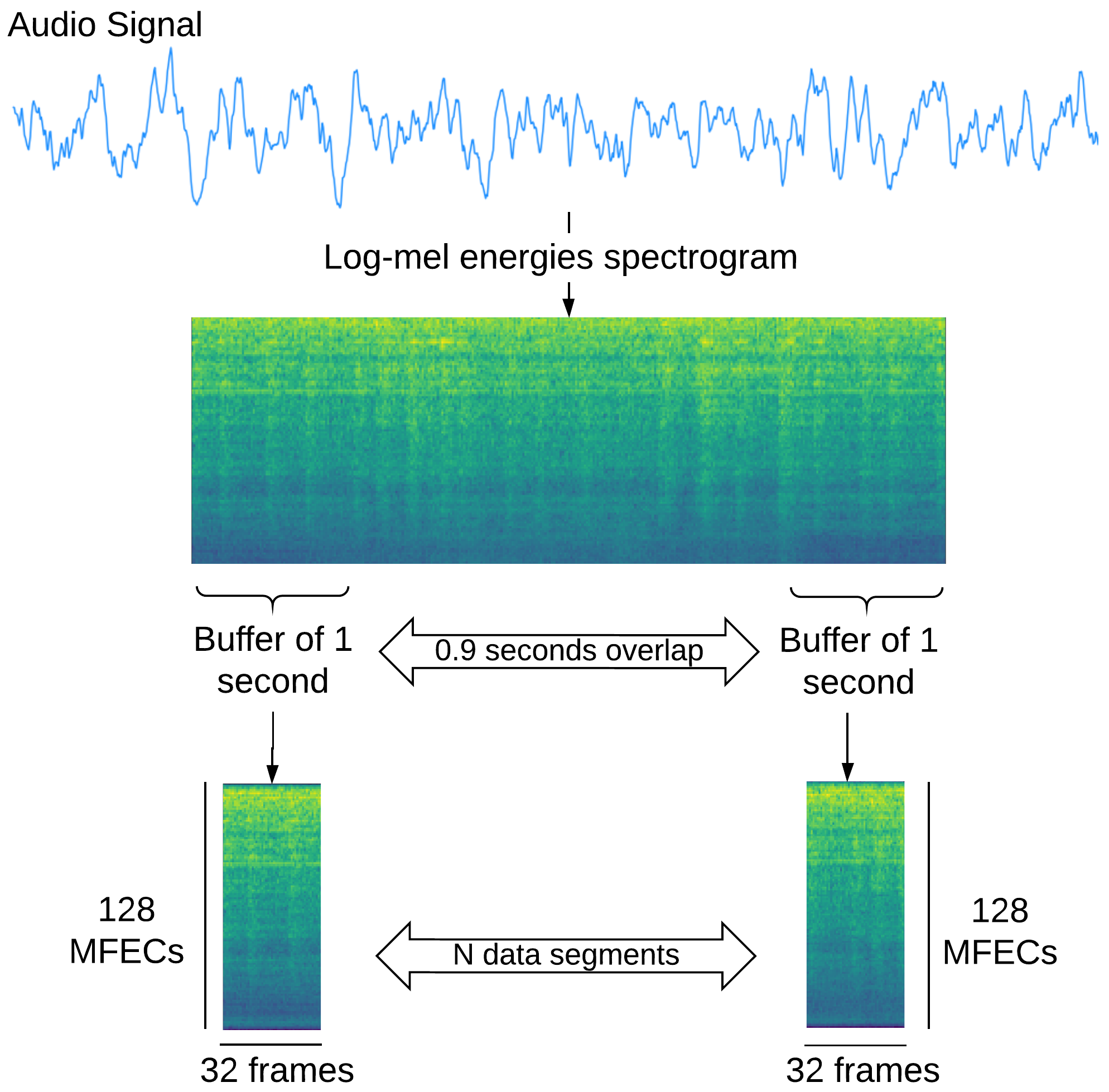}}
      \caption{Feature extraction procedure of the CNN AE.}
      \label{fig:features_conv}
    \end{figure}

\subsection{Training Settings}
\label{ssec:procedure}

The encoder and decoder were trained to minimize the Mean Squared Error (MSE) between input and its reconstruction.
Both architectures were trained with a learning rate of $0.001$ and the Adam optimizer.
The training process is stopped early when the validation loss has stopped improving
for $10$ epochs and the best saved model is selected.
The training procedure was iterated up to a maximum of $100$ epochs.
The batch size for the Dense AE and Convolutional AE algorithms was set as 512 and 64, respectively.

\subsection{Results and Discussion}

All developed architectures were implemented in Python programming language and experiments were conducted in several GPUs (Titan Xp and 1080Ti).
To evaluate the model performance, both AUC and pAUC metrics were used, as defined in the task description \cite{koizumi2020}.
Table \ref{tab:results} presents the AUC and pAUC values for each specific machine and mean values for each machine type, 
obtained for the development dataset by both AE models. For comparison purposes, the baseline system results are also provided in the table.
    \begin{table*}[!htb]
    \caption{Performance results for DCASE 2020 Task 2 for the development dataset (best mean values in \textbf{bold})}  
    \centering
    \label{tab:results}
    \begin{tabular}{lccclcclcc}
    \hline
    \multirow{2}{*}{Machine Type} & \multirow{2}{*}{Machine ID} & \multicolumn{2}{c}{Baseline} & \multicolumn{1}{c}{} & \multicolumn{2}{c}{Dense AE} & \multicolumn{1}{c}{} & \multicolumn{2}{c}{Conv AE} \\ \cline{3-4} \cline{6-7} \cline{9-10}
     &  & \multicolumn{1}{c}{AUC $(\%)$} & \multicolumn{1}{c}{pAUC $(\%)$} & \multicolumn{1}{c}{} & \multicolumn{1}{c}{AUC $(\%)$} & \multicolumn{1}{c}{pAUC $(\%)$} & \multicolumn{1}{c}{} & \multicolumn{1}{c}{AUC $(\%)$} & \multicolumn{1}{c}{pAUC $(\%)$} \\ \hline
    \multirow{5}{*}{ToyCar} & 1 & 81.36 & 68.40 &  & 83.87 & 72.64 &  & 81.59 & 71.88 \\
     & 2 & 85.97 & 77.72 &  & 87.56  & 80.35 &  & 85.46  & 79.92 \\
     & 3 & 63.30 & 55.21 &  & 63.12	 & 55.02 &  & 62.73 & 55.08 \\
     & 4 & 84.45 & 68.97 &  & 88.60 & 	76.68  &  & 82.38 & 69.60 \\ \cline{2-10}
     & \textbf{Average} & 78.77 & 67.58 &  & \textbf{80.79} & \textbf{71.17} &  & 78.04 & 69.12 \\ \hline
    \multirow{4}{*}{ToyConveyor} & 1 & 78.07 & 64.25 &  & 81.67 & 69.41 &  & 79.90 & 62.71 \\
     & 2 & 64.16 & 56.01 &  & 68.04 & 58.31 &  & 67.78 & 54.85 \\
     & 3 & 75.35 & 61.03 &  & 79.59 & 63.64 &  & 80.11 & 62.53 \\ \cline{2-10}
     & \textbf{Average} & 72.53 & 60.43 &  & \textbf{76.43} & \textbf{63.79} &  & 75.93 & 60.03 \\ \hline
    \multirow{5}{*}{fan} & 0 & 54.41 & 49.37 &  & 56.73 & 49.72 &  & 51.77 & 49.05 \\
     & 2 & 73.40 & 54.81 &  & 79.60  & 54.00 &  & 72.71 & 55.51 \\
     & 4 & 61.61 & 53.26 &  & 70.11 &  54.11 &  & 62.60 & 52.80 \\
     & 6 & 73.92 & 52.35 &  & 81.69 & 55.15 &  & 80.05 & 53.19 \\ \cline{2-10}
     & \textbf{Average} & 65.83 & 52.45 &  &  \textbf{72.03} &  \textbf{53.25} &  & 66.78 & 52.63 \\ \hline
    \multirow{5}{*}{pump} & 0 & 67.15 & 56.74 &  & 66.94 & 56.83 &  & 66.37 & 54.95 \\
     & 2 & 61.53 & 58.10 &  & 60.77 &  60.31&  & 54.31 & 53.58 \\
     & 4 & 88.33 & 67.10 &  & 87.00 & 66.32 &  & 94.64 & 77.26 \\
     & 6 & 74.55 & 58.02 &  & 77.53 & 60.32 &  & 76.97 & 58.05 \\ \cline{2-10}
     & \textbf{Average} & 72.89 & 59.99 &  & \textbf{73.06} & 60.94 &  & 72.07  & \textbf{60.96}  \\ \hline
    \multirow{5}{*}{slider} & 0 & 96.19 & 81.44 &  &  96.12 & 82.30 &  & 98.86 & 94.47 \\
     & 2 & 78.97 & 63.68 &  & 79.55 & 64.42 &  & 84.06 & 69.33 \\
     & 4 & 94.30 & 71.98 &  & 95.44 &  76.14 &  & 97.69 & 87.82 \\
     & 6 & 69.59 & 49.02 &  & 77.22 & 49.56 &  & 86.46 &  53.16\\ \cline{2-10}
     & \textbf{Average} & 84.76 & 66.53 &  & 87.08 & 68.10 &  & \textbf{91.77} & \textbf{76.20} \\ \hline
    \multirow{5}{*}{valve} & 0 & 68.76 & 51.70 &  & 74.61 & 52.28 &  & 78.69 & 52.59 \\
     & 2 & 68.18 & 51.83 &  & 76.68 & 52.72 &  & 85.02 & 55.92 \\
     & 4 & 74.30 & 51.97 &  & 79.58 & 50.96 &  & 82.59 & 53.68 \\
     & 6 & 53.90 & 48.43 &  & 57.78 & 48.73 &  &  69.03 & 50.22 \\ \cline{2-10}
     & \textbf{Average} & 66.28 & 50.98 &  & 72.16 & 51.17 &  & \textbf{78.83} & \textbf{53.10} \\ \hline
    \end{tabular}
    \end{table*}

In terms of mean AUC and pAUC values for each machine type, the Dense AE outperforms the baseline system in every machine type.
Furthermore, the baseline system only achieved better results in a few specific machines, namely ToyCar ID 3, pump IDs 0, 2 and 4, and slider ID 0.
Regarding the CNN AE, in general the model outperformed the baseline system, 
although the latter achieved higher mean AUC values for 2 of 6 machine types (ToyCar and pump).
The two proposed AE are quite competitive in terms of mean AUC and pAUC values, with CNN AE outperforming Dense AE only in 2 of 6 machine types 
(slider and valve).
Overall, both the dense and CNN AE outperform the baseline system in both anomaly detection metrics (AUC and pAUC).
Considering that none of the proposed AE models obtained the best results for all machine types, we have created
a third method for the competition, which is termed mixed approach. This third approach uses the best AE for each machine type,
namely the CNN AE is used for the slider and valve machines, while the Dense AE is adopted for the other machine types.
All the developed code is available on github\footnotemark.

\footnotetext{\url{https://github.com/APILASTRI/DCASE_Task2_UMINHO}}

\section{CONCLUSIONS}
\label{sec:conclusion}

In this paper, we proposed two Autoencoder (AE) models for an unsupervised Anomalous Sound Detection (ASD), 
for the second task of the DCASE 2020 challenge.
The AE models are based on Dense and Convolutional Neural Networks (CNN). Several preliminary experiments were conducted, 
resulting in two proposed AE architectures that use sound energy features from mel-spectograms.
Using the provided challenge datasets, the two deep AE were trained and tested with the competition six machine types.
Overall, competitive results were obtained when compared with the challenge baseline model.
For two machine types (slider and valve), the best results were achieved by the CNN AE, while the Dense AE provided the best results for the other machines
(ToyCar, ToyConveyor, fan and pump). Thus, a third method was proposed for the competition, which uses the best AE model for each machine type. 
We consider that the obtained AE results are of quality. For instance, the achieved test data AUC values range from 72\% (good) to  92\% (excellent discrimination).

As future work, we aim to explore with more depth the proposed AE structures. For instance, by adopting audio data augmentation techniques 
(e.g., pitching, time-shifting) to improve the training results.
Furthermore, we intend to explore other neural network architectures for sound anomaly detection, such as Generative Adversarial Networks (GAN) 
and Variational AEs.

\section{ACKNOWLEDGMENTS}
\label{sec:ack}

This work is supported by the European Structural and Investment Funds in the FEDER component, through the Operational Competitiveness and Internationalization Programme (COMPETE 2020) - Project nº 039334; Funding Reference: POCI-01-0247-FEDER-039334.


\bibliographystyle{IEEEtran}

\bibliography{dcase2}

\end{sloppy}
\end{document}